\documentstyle[epsfig]{article}

\begin{document}

%   \hyphenation{Ma-sa-ryk}
%   \hyphenation{Kho-lo-pov}
%   \hyphenation{Ha-dra-va}
%   \hyphenation{Mi-ku-l\'{a}-\v{s}ek}
%   \hyphenation{On-d\v{r}e-jov}

\title{Photometric study of eclipsing binary RV Tri}

\author{F. Hroch, M. Ba\v{s}n\'{y}}

%\institute{Institute of Theoretical Physics and Astrophysics,
%           Faculty of Science, Masaryk University,\\
%           Kotl\'{a}\v{r}sk\'{a} 2, 611 37 Brno, Czech Republic, 
%           email: hroch@physics.muni.cz, mbasny@physics.muni.cz}

\maketitle

\begin{abstract}
   We present new photometric observations of RV~Tri. 
   A complete light curve was obtained in the R and V filters.  
   These light curves were processed by the Wilson-Devinney (WD) code
   and the FOTEL to get solution of this eclipsing binary system.
   We discuss characteristics of the system
   and compare our model with a previous calculation. 
   Our observations of primary minima and the comparison with
   those predicted on the base of the GCVS catalogue indicates
   that there are 
   significant differences.  We discuss these differences in 
   the context of period change. 

%   \keywords{photometry --  eclipsing binary} 
              
\end{abstract}

%__________________________________________________________________
\section{Introduction}

RV~Tri (GSC 2321.0070) was recognised as a variable star by W. Strohmeier 
and the discovery was published in Geyer~et~al.~(\cite{Geyer}), where
preliminary values of the light elements were given.  
RV~Tri is classified as an Algol star type,
the depth of primary and secondary minima being 1.2 and 0.1 in magnitude, 
respectively. The times of primary minima have been frequently visually 
determined during last three decades, but a detailed analysis 
of this star is still missing. Therefore, we have observed RV~Tri  
with a new CCD camera equipment at our 
observatory in the summer of 1997. The mean light curve which we obtained from 
the observations enables us to study this system. 

%-----------------------------------------------------------------------
\section{Observations}
The data presented in this paper were obtained during
the summer/autumn of 1997 and 1998. 
We used the 0.60~m reflector of our observatory. 
The CCD front side camera placed
at the Newtonian focus took 30 or 60 seconds exposures of the variable 
and comparison stars. V (Johnson) and R (Kron-Cousins) standard filters were 
used.
The observations are summarised in Table~\ref{tabObs}.

\begin{table*}
\caption{The list of observations and the primary minima of RV~Tri 
         (the (O-C) differences from the GCVS ephemeris as mean of R,V).}
\label{tabObs}
\begin{flushleft}
%\begin{tabular}{llllll}
\begin{tabular}{llll}
\hline
night             & minimum in R    & minimum in V     
                  %& No. R+V    & epoch 
                  & (O-C) \\
\hline
1997 -- 09 -- 20/21 & ---         & ---         
                  %  & 141          &  {} 
                  & {} \\ 
1997 -- 09 -- 21/22 & ---         & ---         
                  %  & 95         & {}     
                  & {} \\
1997 -- 09 -- 28/29 & 50720.3480 $\pm$ 0.0057 
                    & 50720.3470 $\pm$ 0.0057 
                  %  & 179          & 6219  
                  & $-0.0120 \pm 0.0081$ \\
1997 -- 10 -- 19/20 & 50741.4525 $\pm$ 0.0096
                    & 50741.4517 $\pm$ 0.0103 
                  %  & 85          & 6247   
                  & $-0.0085 \pm 0.0141$ \\
1997 -- 10 -- 31/11 -- 01 
                    & 50753.5083 $\pm$ 0.0036
                    & 50753.5074 $\pm$ 0.0038 
                  %  & 104          & 6263  
                  & $-0.0156 \pm  0.0052$ \\
1997 -- 11 -- 10/11 & 50763.3054 $\pm$ 0.0065
                    & 50763.3061 $\pm$ 0.0058 
                  %  & 54          & 6276   
                    & $-0.0146 \pm 0.0087$ \\
1998 -- 09 -- 24/25 & 51081.3521 $\pm$ 0.0032
                    & 51081.3532 $\pm$ 0.0038
                    %& 255         & 6698   
                    & $-0.0145 \pm 0.0050$ \\
1998 -- 11 -- 11/12 & 51130.3420 $\pm$ 0.0035
                    & 51130.3422 $\pm$ 0.0063 
                    %& 148         & 6763   
                    & $-0.0134 \pm 0.0072$ \\
\hline
\end{tabular}
\end{flushleft}
\end{table*}

%------------------------------------------------------------------
   \begin{figure}[tb]
\vspace{0cm}
\hspace{0cm}\makebox{\psfig{figure=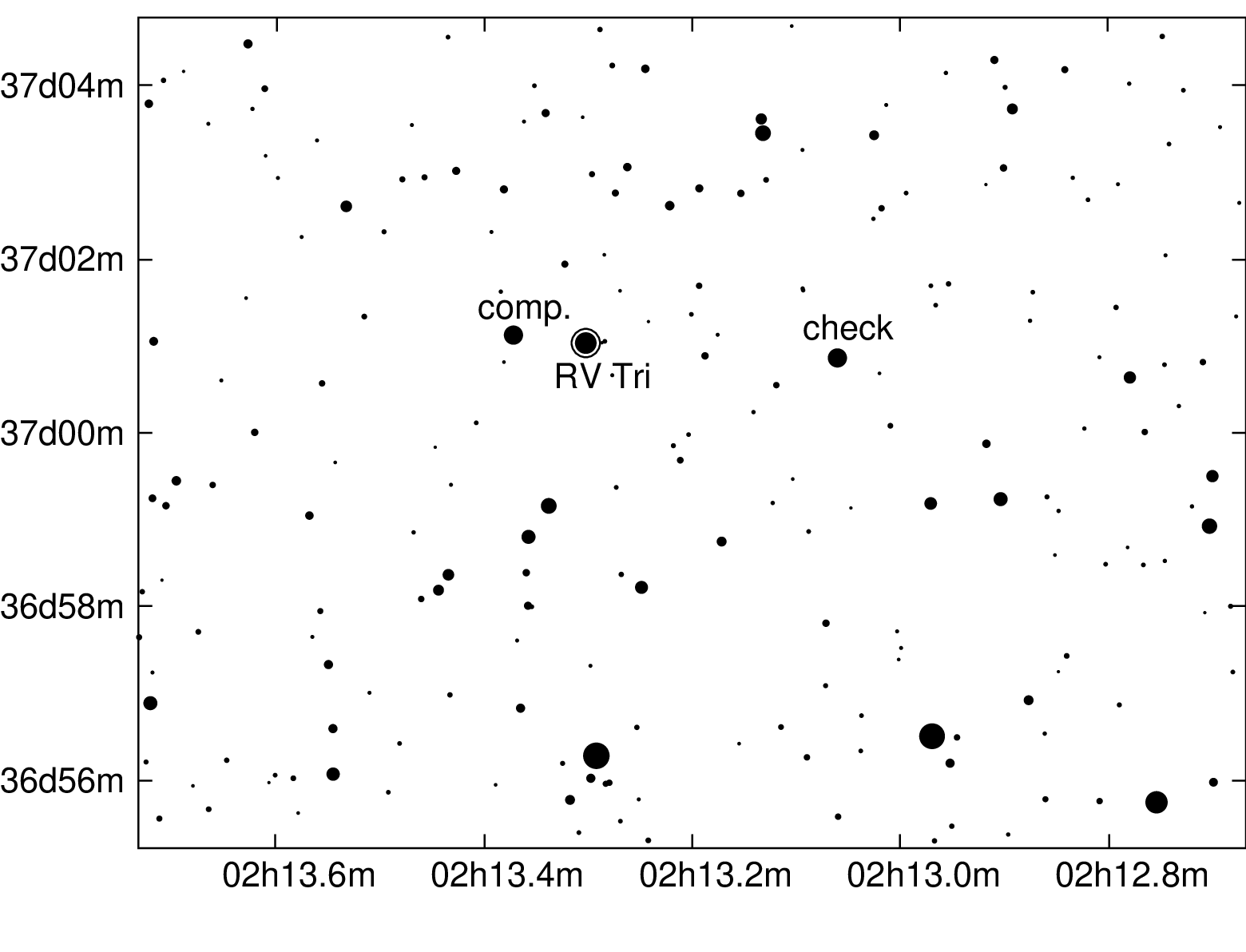,width=8.0cm}}
\vspace{0cm}
%      \vspace{5cm}
\caption[]{Identification chart constructed on the base of the USNO catalogue
           for RV~Tri (eq. J2000.0).}
         \label{obrMap}
   \end{figure}
%------------------------------------------------------------------

The transformation between our instrumental 
magnitudes and Johnson photometric system gives
as extra-atmospheric magnitudes and colour indexes of the comparison
and check star the values listed in Table~\ref{tabcomp}.
We used Landolt photometric standards (Landolt~(\cite{Landolt})) for 
derivation of these transformations. 

The standard image processing was applied to every image. 
The synthetic differential aperture photometry was done
by MUNIPHOT (Hroch~(\cite{Hroch})). 
Differential photometry of both variable and comparison 
stars were not reduced for the extinction because the stars in question
are very close each to other in the sky ($\sim 1'$). The chart of the 
close environment of RV~Tri with comparison and check star is shown in 
Fig.~\ref{obrMap}. Its precise position as determined by the processing 
of our observations (J2000.0, the proper motion neglected) is:
\begin{displaymath}
\begin{array}{ll}
         \alpha = 2^{\rm h} 13^{\rm m} 18^{\rm s}\!\!.12 
       & \delta = + 37^{\circ} 01' 02''\!\!.1 \\
\end{array}
\end{displaymath}
The position error was smaller than $0''\!\!.2$ for both coordinates.
We choose the USNO-A catalogue as astrometric reference catalogue. 

%-------------------------------------------------------------------
\begin{table}
\caption{The extra-atmospheric magnitudes and colour index 
         of observed stars in Johnson photometric system.}
\label{tabcomp}
\begin{flushleft}
\begin{tabular}{lll}
\hline
star    & V                  & (V - R) \\
\hline
comp. (GSC 2321.1715)  & 13.096 $\pm$ 0.089 & 0.387 $\pm$ 0.094 \\
check (GSC 2321.0692)  & 13.349 $\pm$ 0.082 & 0.713 $\pm$ 0.093 \\
RV Tri, max.  &  11.0      & 0.2     \\
RV Tri, min.  &  12.2      & 0.3     \\
\hline
\end{tabular}
\end{flushleft}
\end{table}
%------------------------------------------------------------------

\section{Light elements} \label{elements}

%------------------------------------------------------------------
\begin{figure}[bt]
\begin{center}
\input{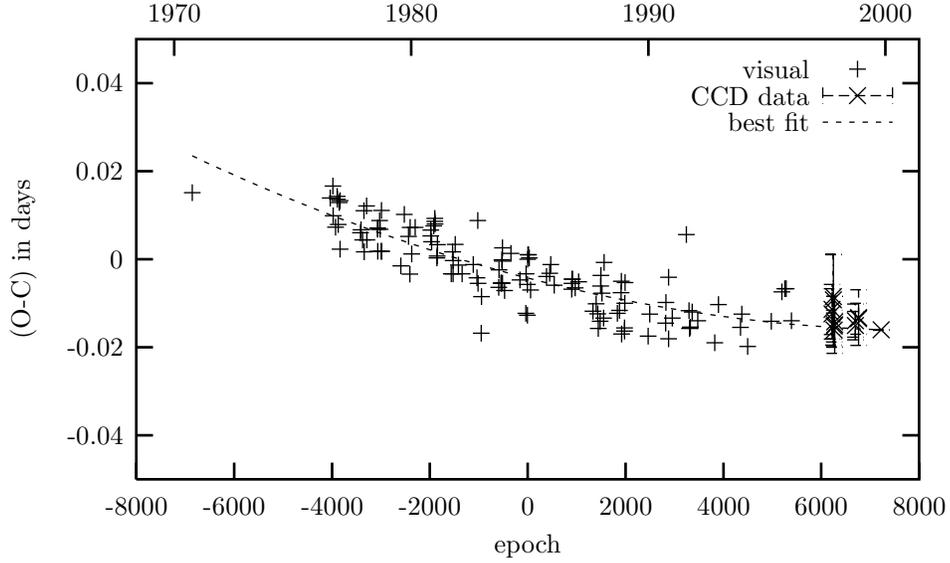}
\end{center}
%\vspace{0cm}
%\hspace{0cm}\psfig{figure=obroc.ps,width=8.0cm}
%\hspace{0cm}\psfig{figure=fit.ps,width=8.0cm}
%\vspace{0cm}

%      \vspace{5cm}
\caption[]{The O--C diagram for RV~Tri. The GCVS light elements
           were used. %The line corresponds to the elements (\ref{vizline}).
           The plus '+' represents minima 
           reported by visual observers.
% over the world 
%           without Czech and Slovak observers, their data are 
%           labelled by the crosses '$\times$'. 
           We estimated errors
           of visually determined minima on 10 minutes.
%           The asterisk '$\ast$' with errorbar on right edge 
           The six points with error-bars on the right edge
           of the picture represents our minima. The last point is 
           minimum from Nelson~(\cite{nelson}).}
         \label{obrviz}
   \end{figure}
%------------------------------------------------------------------

The values of the light elements published in the original
paper by Geyer et. al.~(\cite{Geyer}) and values of the elements given 
by the last edition of the GCVS catalogue %\footnote{
(Kholopov~et~al.~(\cite{Kholopov})) witch ephemeris
\begin{displaymath} 
\mathrm{Min \, I (J.D. Hel.)} =
       24 46 033^{\mathrm{d}}\!\!.308 + 0^{\mathrm{d}}\!\!.75366648 E 
\end{displaymath}
are highly 
inaccurate in comparison with the accuracy of the time of the minima 
determinations derived from our observations. We have observed an (O--C) 
difference of about 15 minutes at current epochs. 
This fact is also illustrated 
by the (O--C) diagram compiled from visual observations of the 
primary minima times  
of RV~Tri (see Fig.~\ref{obrviz}) currently entered 
in the B.R.N.O. database (Variable Star Section of Czech Astronomical
Society). 

%The procession
%of all times of visual minima gives us improved light elements (computed
%by the robust linear least square method only for 152 accepted 
%points)
%\begin{equation} \label{vizline}
%{\rm  Min \, I (J.D. Hel.)} =
%  \begin{array}[t]{r}
%                 24 46 033^{\rm d}\!\!.3051 \\ 
%                     \pm 0^{\rm d}\!\!.0004 \\
%  \end{array} +
%  \begin{array}[t]{r}
%                   0^{\rm d}\!\!.75366361 \\  
%               \pm 0^{\rm d}\!\!.00000015
%  \end{array} E    
%\end{equation}  
We have tried to verify this systematic trend in the light elements
of RV Tri
% these elements
 by observation of some primary minima
during our observational periods. The times of the minima for each 
observational set were computed by the Kwee and van Woerden method 
(\cite{Kwee}). Our and Nelson's~(\cite{nelson}) observations 
(see Fig.~\ref{obrviz}) 
confirm this behaviour.
The procession of all times of visual minima gives us improved 
quadratic ephemeris (computed by the weighted robust linear least 
square method only for 152 accepted points)
\begin{equation} \label{ourel}
{\rm  Min \, I (J.D. Hel.)} =
  \begin{array}[t]{r}
                 24 46 033^{\rm d}\!\!.30373 \\ 
                     \pm 0^{\rm d}\!\!.00049 \\
  \end{array} +
  \begin{array}[t]{r}
                   0^{\rm d}\!\!.75366361 \\ 
               \pm 0^{\rm d}\!\!.00000015
  \end{array} E +   
  \begin{array}[t]{r}
                   1^{\rm d}\!\!.72 \times 10^{-10}  \\ 
               \pm 0^{\rm d}\!\!.25 \times 10^{-10}
  \end{array} E^2    
\end{equation}  
which approximates the time of the minima in the range  between JD = 24~50~720
and JD = 24~51~130 with a precision better than $\pm$0.004 days.
The use of this ephemeris out of this range is uncertain.
% do not satisfy the elements (\ref{vizline}) with required 
%precision (differences about 10 minutes), 
We used this ephemeris
%(the same period, but the basic minimum a mid our minima)
for reduction of our observations to the mean light curve.
We estimated the error of visually determined minima to be 10 minutes.
%\begin{equation} \label{ourel}
%{\rm  Min \, I (J.D. Hel.)} =
%  \begin{array}[t]{r}
%		  24 50 924^{\rm d}\!\!.5898 \\ 
%		      \pm 0^{\rm d}\!\!.0003 \\
%  \end{array} +
%  \begin{array}[t]{r}
%		    0^{\rm d}\!\!.75366354 \\ 
%		\pm 0^{\rm d}\!\!.00000015
%  \end{array} E    
%\end{equation}  
%These elements approximate times of minima in range  between JD = 2450720
%and JD = 2451130 with precision better than $\pm$0.0004 day.
%The use of the above elements out of this range is uncertain.

%__________________________________________________________________

\section{Light curve analysis}

We employed the observational data of RV~Tri for the acquisition of
orbital and physical parameters of the system.
The results of 530 measurements in the R and 530 measurements 
in the V filter have been processed by the programs WD98 and FOTEL. 

%\begin{table}[bt]
%\caption{List of parameters.}
%\label{tabpar}
%\begin{flushleft}
%   \[
%      \begin{array}{lp{0.8\linewidth}}
%	 \hline
%	 {\rm parameter} & description \\
%	 \hline
%	  i   \dotfill & orbital inclination \\
%	  l_1 \dotfill & relative luminosity, primary\\
%	  l_2 \dotfill & relative luminosity, secondary\\
%	  l_3 \dotfill & relative luminosity, third light \\
%	  r_1 \dotfill & relative stellar radius, primary\\
%	  r_2 \dotfill & relative stellar radius, secondary\\
%	  x_1 \dotfill & limb darkening of primary component\\
%	  x_2 \dotfill & limb darkening of secondary component\\
%	  g_1 \dotfill & gravity darkening of primary component\\
%	  g_2 \dotfill & gravity darkening of secondary component\\
%	  A_1 \dotfill & albedo of the primary\\
%	  A_2 \dotfill & albedo of the secondary\\
%	  T_1 \dotfill & effective surface temperature, primary \\
%	  T_2 \dotfill & effective surface temperature, secondary \\
%	  q   \dotfill & mass ratio $m_2/m_1$\\
%	  F_1 \dotfill & synchronisation index, primary \\
%	  F_2 \dotfill & synchronisation index, secondary \\
%	 \hline
%      \end{array}
%   \]   
%\end{flushleft}
%\end{table}

\subsection{Wilson-Devinney}
\label{wd} 
WD98 --- was written by R.E.Wilson and E.J.Devinney
described in Wilson, Devinney~(\cite{wilson71}).
We uses more recent WD98  modification. An operation mode
for Algol type stars was run.

\subsection{FOTEL}
\label{FOTEL}
FOTEL --- a computer program for the determination of {\em phot\/}ometric
{\em el\/}ements --- was written by P.~Hadrava for data processing
at the Ond\v{r}ejov observatory (Hadrava~(\cite{Hadrava})). FOTEL 
simultaneously calculates the light curve and the radial velocities with the
aim to determine the parameters of the studied system. The latest 
version (1994) approximates the Roche potential by an effective potential.
%shape of stars in
The construction of a binary system model follows  
the paper by Wilson~(\cite{Wilson}), but many improvements made 
by the author are included.

\subsection{Data analysis} \label{analysis}

The shape of the light curve of RV Tri is similar to light curves
of Algol-like systems. Therefore, we supposed that the system 
is semi-detached. Some parameters of a semi-detached 
system are mutually dependent. A radius of the secondary (filling component)
$r_2$ and the mass ratio
$q = M_2/M_1$ satisfies an approximative condition developed by
Eggleton~(\cite{Eggleton})
%\begin{equation} \label{l1}
%r_2 = \frac{0.49 \, q^{2/3}}{0.6 \, q^{2/3} + \ln(1 + q^{1/3})} 
%\end{equation}
%(for all $q$) 
and reduces the number of free parameters. Note, the $r_2$
(and $r_1$, of course) depends strongly on the shape of the light curve,
but the mass ratio is poorly determined from the photometric 
observations. 
  
When the system is at maximum light, the star surfaces are not
obscured and the effective temperature of whole system can be
estimated %determined as great then the temperature derived 
from the colour index (V--R). 
The colour index of the system at light maximum (0.2 mag) 
indicates that the system contains a star of A (or hotter)
spectral type. 

On the other hand, when the larger part of the luminous component 
is obscured, the system is at minimum light and it is more red 
with a  (V--R) index of 0.3~magnitude. This colour change implicates, 
that the secondary component is of F (or colder) spectral type. 
We have used tables by Allen~(\cite{Allen}) to estimate  
the relation between the colour index and the spectral type.  
%The colour change is a small, we expects also that luminosities
%of both stars are a very different (10:1 at least).

This raw colour index analysis and a well--known 
characteristic of the Algol-like systems leads to the conclusion 
that the observed light changes consists 
of a semi-detached binary system, which contains a hot main-sequence 
primary and a cold sub-giant secondary star. The secondary
star probably fills its Roche lobe. We suppose in our initial computations 
that the mass of the secondary is only half of that of the primary  
%The preliminary light curve analysis confirms our hypothesis. 
%The secondary star 
and that it is highly-evolved, in spite of being less massive, according
to the Algol paradox.

These results of the preliminary analysis of the light curve
provided us with the basic parameters which then served as input data for
processing with the programs WD98 or FOTEL. The values of the
second order parameters (limb darkening, albedo, gravity darkening)
was used from tables by Claret~(\cite{claret}).
The use third light leads to residues not significantly different
from no-third light solution. The value of the third light
parameter was comparable with its statistical and our data measure error.
%We attempted to check 
%accuracy of the previous calculations and to precise them, if needed. 
%The calculations were realized simultaneously in the R and the V filters.

The  resulting physical parameters are given in 
Table~\ref{maintab}. The WD calculated light curve of the system RV~Tri 
together with the observational data in the R filter are presented in 
Fig.~\ref{lcRVTriR}. The (O--C) residuals  were also drawn (on the top of 
Fig.~\ref{lcRVTriR}) for better appreciation of the calculation.
 The light curve, the observational data and the residuals
for the V filter are presented in Fig.~\ref{lcRVTriV}.

\begin{table}
\caption{Photometry solutions of RV Tri.}
\label{maintab}
\begin{center}
\begin{displaymath}
\begin{array}{lllll}
\hline
{} & \multicolumn{2}{c}{\mathrm{WD98}} & \multicolumn{2}{c}{\mathrm{FOTEL}} \\
{} & \mathrm{V} & \mathrm{R}           & \mathrm{V} & \mathrm{R}  \\
\hline
i        & 84^{\circ}\!\!.5\pm1^{\circ}\!\!.0
	 & {}
         & 82^{\circ}\!\!.4\pm1^{\circ}\!\!.0
	 & {}\\
l_1      & 0.92\pm0.14
         & 0.94\pm0.59
         & 0.91
	 & {} \\
l_2      & 0.08
         & 0.06
         & 0.09
	 & {} \\
x_1^*    & 0.51
	 & 0.43
	 & 0.51
         & 0.43 \\
x_2^*    & 0.83
	 & 0.74
	 & 0.83
         & 0.74 \\
g_1^*    & 1.0
         & {}
         & 1.0
         & {} \\
g_2^*    & 0.4
         & {}
         & 0.4
         & {} \\
\Omega_1 & 3.41\pm0.10
          & {}
          & {} 
          & {} \\
\Omega_2 & 2.6
          & {}
          & {} 
          & {} \\
A_1^*     & 1.0
          & {}
          & 1.0
          & {} \\
A_2^*     & 0.5
          & {}
          & 0.5
          & {} \\
T_1 (\mathrm{K}) & 9500\pm300
                 & {}
                 & 9500\pm1400 
                 & {}\\
T_2 (\mathrm{K}) & 4900
                 & {}
                 & 5300\pm200
                 & {} \\
q                & 0.34\pm0.03
                 & {}
                 & 0.34\pm0.41
                 & {}\\
r_1(\mathrm{pole}) & 0.32\pm0.01
                   & {}
                   & 0.33
                   & {}\\
r_1(\mathrm{point}) & 0.33\pm0.01
                   & {}
                   & 0.33
                   & {}\\
r_1(\mathrm{side}) & 0.33\pm0.01
                   & {}
                   & 0.34
                   & {} \\
r_1(\mathrm{back}) & 0.33\pm0.01
                   & {}
                   & 0.33
                   & {} \\
r_2(\mathrm{pole}) & 0.29\pm0.01
                   & {}
                   & 0.30
                   & {} \\
r_2(\mathrm{point}) & 0.40\pm0.05
                   & {}
                   & 0.39
                   & {} \\
r_2(\mathrm{side}) & 0.29\pm0.01
                   & {}
                   & 0.33
                   & {} \\
r_2(\mathrm{back}) & 0.32\pm0.01
                   & {}
                   & 0.32
                   & {}\\
\hline
\end{array}
\end{displaymath}
\end{center}
* {\rm not fitted}, 
  $i$ - orbital inclination,
  $l_1,l_2$ - relative luminosity,
  $x_1, x_2, g_1, g_2$ - limb and gravity darkening,
  $A_1, A_2$ - albedo,
  $T_1, T_2$ - effective surface temperature,
  $\Omega_1, \Omega_2$ - potentials,
  $q$ - mass ratio $m_2/m_1$,
  $r_1, r_2$ - relative radii
%   $i$ - orbital inclination, 
%   $l_1$ - relative luminosity of the primary,
%   $l_2$ - relative luminosity of the secondary,
%   $l_3$ - relative luminosity of the third light,
%   $r_1$ - relative stellar radius of the primary,
%   $r_2$ - relative stellar radius of the secondary,
%   $x_1$ - limb darkening of primary component,
%   $x_2$ - limb darkening of secondary component,
%   $g_1$ - gravity darkening of primary component,
%   $g_2$ - gravity darkening of secondary component,
%   $A_1$ - albedo of the primary,
%   $A_2$ - albedo of the secondary,
%   $T_1$ - effective surface temperature of the primary,
%   $T_2$ - effective surface temperature of the secondary,
%   $q$ - mass ratio $m_2/m_1$,
%   $F_1$ - synchronisation index of the primary,
%   $F_2$ - synchronisation index of the secondary.
\end{table} 
%%%%%%%%%%%%%%%%%%%%%%%%%%%%%%%%%%%%%%%%%%%%%%%%%%%%%%%%%%%%%%%%%%%%%%%%%%

%
%On the basis of the mass ratio of the components $q=1.0$,  
%the shape of the Roche lobe was calculated (Fig.~\ref{ekRVTri}).
%The calculation of the radii of the individual components
%($R_1 \sim 0.35$,  $R_2 \sim 0.31$) showed that none
%of the components is filling the Roche lobe. 
%
%The  resultant astrophysical 
%parameters with their statistical errors are adduced in 
%Table~\ref{maintab}.
%The calculated light curve of the system RV~Tri together
%with the observation data in R filter are presented in 
%Fig.~\ref{lcRVTriR}.  For better appreciation of the
%calculation the residues (O--C) 
%were also drawn (the top of Fig.~\ref{lcRVTriR}).
%The light curve, the observation data and residues for V filter are 
%presented in Fig.~\ref{lcRVTriV}. 
%In the table we see that the  EBOP solutions obtained for R and V filters 
%are within the errors the same. 
%The result would probably be even more accurate if we could fit both
%the curves together in EBOP case.

%%%%%%%%%%%%%%%%%%%%%%%%%%%%%%%%%%%%%%%%%%%%%%%%%%%%%%%%%%%%%%%%%%%%%%%%%%
   \begin{figure}
     \begin{center}
      \input{r.pslatex}
     \end{center}	
      \caption{R filter data and model fit of
               light curve of RV~Tri.}
      \label{lcRVTriR}
%   \end{figure}
   
%   \begin{figure}[tb]
     \begin{center}
      \input{v.pslatex}
     \end{center}	
      \caption{V filter data and model fit  of 
               light curve of RV~Tri.}
      \label{lcRVTriV}
   \end{figure}
%%%%%%%%%%%%%%%%%%%%%%%%%%%%%%%%%%%%%%%%%%%%%%%%%%%%%%%%%%%%%%%%%%%%%%%%%

\section{Conclusions}

The accuracy of the GCVS light elements of the RV~Tri is significantly
lower than that of our minima determination. Therefore, we derived
improved elements using visual observations
obtained during the last 30 years (see Sect~\ref{elements})
and a few of our measurements. The credibility of the visual observations 
is sufficient in this case, because the amplitude of the light 
changes is greater than one magnitude. The observed 
primary minima are good described by the quadratic fit of the data.
%The elements derived in this way do not 
%correspond to our minima observations entirely. 
%The extrapolation of elements (\ref{vizline}) to the epoch $E=6500$ 
%predicts (O-C) difference of $-0.022\pm0.005$ days, but the mean of our
%observations indicates difference of only $-0.014\pm0.003$ days. 
%This significant difference 
Its behaviour probably does not have origin in statistical
errors but it is caused by period change during the latest epoch.
These observations of all primary minima indicates a change
of light elements, which is probably caused due to  
slight real changes in RV~Tri orbital parameters.
This hypothesis can be additionally tested by the construction of 
a physical model of this binary system.

%. These results correspond to the elements 
%determined from the visual 
%minima  (eq. (\ref{vizline})) and to
%the elements independently found in \cite{SAC}. 
% The determination of the 
%visual elements by (\ref{vizline})? is probably the most precise
%one concerning these parameters of RV~Tri.

We have used two methods for the analysis of the observed light curve 
of RV~Tri  (computer programs WD98 and FOTEL). 
%The constructed model 
%approximates the observed data with satisfactory precision.  
The model approximation and the results of both programs
are consistent within statistical errors and different processing
methods.
% in spite of the fact that FOTEL uses a more realistic model. 
The radial velocities of the components are required
for the complete analysis of this system, but they 
are not available for RV~Tri yet.

The derived solution can include systematic errors due to 
the small sensitivity of the photometric data on some parameters
(especially on $q$). Therefore, all of the parameters 
($i$, $r_1$, $T_1$, $T_2$) are strongly correlated to $q$.
Note that $r_1$ and $r_2$ are bounded by Eggleton's relation.
This correlation and the change of $q$ from $0.4$ to $0.5$ produces  
different values of parameters as compared to the results given by 
Brancewicz and Dworak~(\cite{Brancewicz})
The increase of $q$ induce a smaller value of $r_1, r_2$ 
and the decrease of the stars' effective temperatures $T_1, T_2$. 
The whole luminosity of the system is conserved in this way.    

We deduce from the best fit of the light curve that the 
binary system RV~Tri is composed of the main-sequence primary  
($T_1 = 9500$ K) and the secondary sub-giant star 
($T_2 = 4900$ K), i.e. of spectral classes near A0, G5.
The stars has been approximated by tri-axial ellipsoids with
the effective radii 0.33 and 0.40 of the separation of their centres. 
The mass ratio has been fitted to $q = 0.36$. We have supposed  
a non-eccentric orbit. The relative luminosity derived by WD98 are 
in good agreement with the FOTEL values: $l_1 = 0.91, l_2 = 0.09$. 

These results partially confirm the model developed by Brancewicz
and Dworak~(\cite{Brancewicz}). These authors introduce the following 
values of the parameters of this system: $q = 0.47$,
$T_1 = 8950$K, $T_2 = 4700$K, $l_1 = 0.89$, $l_2 = 0.11$,
$r_1 = 0.35$, $r_2 = 0.45$ in their paper. 
The disagreements are probably caused by the oversimplified
method or incomplete input data. Note, that their model does not 
satisfy Eggleton's relation (Sect.~\ref{analysis}). They  
supposed a detached binary system in this paper.

We supposed that RV Tri is a semi-detached system in our model.
The sub-giant secondary star fills its Roche lobe. The primary star is
a main-sequence star with 65\% lobe filling. RV Tri is an evolved binary
system after mass exchange. 
The highly evolved secondary star has a smaller mass
than the less evolved primary star. %It oppose with 
This is contrary to star evolution  theories of isolated stars, 
but it is in good agreement
with Algol paradox theories (Wilson~(\cite{Wilson})).   

Figure~\ref{lcRVTriV} shows some distortions
of the observed light curve. The difference between the measured
points and the model light curve does not have only statistical
character, systematic trends can also be clearly seen.  The V light
curve shows an asymmetry, but the equivalent features are strongly 
suppressed in the R filter. The most striking differences are data and fit 
at primary minima and shortly after it. Firstly, we made a hypothesis
about a hot spot on the surface of the primary, which is a deformation 
of the light curve.    
 The careful analysis of our whole observational
equipment gave the conclusion that these asymmetries are due to some
optical inhomogeneities on the surface of the V filter.
This fact influences fitting of the synthetic light
curve. The residuals at primary minima in the V filter presents
distortions of the light curve due to the filter problem and
it is possible that the temperature of the secondary contains
some systematic errors.

%\begin{acknowledgements}
We are very much indebted to R. von Unge, Z.~Mikul\'{a}\v{s}ek, 
J.~Hollan and R.~Hudec for their valuable help and advice. 
Thanks are also due to A.~Paschke, M.~Zejda
and all the visual observers for permission to use their data
from the B.R.N.O. database and P.~Hadrava, R.E.Wilson and E.J.Devinney
for the use of their software. 
We have done much of the computation at the Super-computing 
Centre of the Masaryk University in Brno. 
We have used the USNO catalogue for astrometry and map plotting.
We also wish to thank J.~Brandstettrov\'{a} 
for her valuable comments and reading of the manuscript.
%\end{acknowledgements}

\end{document}